# A design specification for Critical Illness Digital Twins (CIDTs) to cure sepsis: responding to the National Academies of Sciences, Engineering and Medicine Report "Foundational Research Gaps and Future Directions for Digital Twins"

Gary An[1*] and Chase Cockrell[1]
1. Department of Surgery, University of Vermont Larner College of Medicine

*Corresponding Author:
89 Beaumont Ave
Given 319D
Burlington, VT 05405
gan@med.uvm.edu or docgca@gmail.com

## Abstract

On December 15, 2023, The National Academies of Sciences, Engineering and Medicine (NASEM) released a report entitled: "Foundational Research Gaps and Future Directions for Digital Twins." The ostensible purpose of this report was to bring some structure to the burgeoning field of digital twins by providing a working definition and a series of research challenges that need to be addressed to allow this technology to fulfill its full potential. In the work presented herein we focus on five specific findings from the NASEM Report: 1) definition of a Digital Twin, 2) using "fit-for-purpose" guidance, 3) developing novel approaches to Verification, Validation and Uncertainty Quantification (VVUQ) of Digital Twins, 4) incorporating control as an explicit purpose for a Digital Twin and 5) using a Digital Twin to guide data collection and sensor development, and describe how these findings are addressed through the design specifications for a Critical Illness Digital Twin (CIDT) aimed at curing sepsis.

# 1.0 Introduction

## 1.1 Introduction to sepsis

"Sepsis" is a term used to describe the pathophysiological consequences of the acute inflammatory response to severe infection or injury and is functionally similar to a series of other conditions such as "cytokine storm" or cytokine release syndrome, where host tissue damage leading to organ dysfunction arises from a disordered acute immune response and leads to critical illness that requires care in an intensive care setting. While this term is generally applied to the host response to infection, the pathophysiological processes are present following multiple other insults, such as burn injury, trauma, hemorrhagic shock and CAR-T cell therapy, and can be considered nearly ubiquitous in most impactful forms of critical illness. Specifically, sepsis is responsible for more than 215,000 deaths in the United States per year with an annual healthcare cost of over $16 billion [1]. Despite extensive basic science investigations into the mechanisms associated with the disordered acute inflammation, there has been essentially universal failure in the translation of that knowledge into effective therapeutics. While targeted manipulation of various cytokines/mediators, damage-associated molecular pattern molecules, oxygen and nitrogen free radicals, coagulation pathway intermediates, and vasoactive peptides and lipids have shown promise in pre-clinical studies, none of these have been found efficacious in Phase III clinical trials [2, 3]. As such there is currently not a single drug approved by the U.S. Food and Drug Administration that specifically targets the underlying immune pathophysiology of sepsis [2, 4]. Therefore, what we propose is a rational pathway to actually curing sepsis (versus just physiological support), and assert that developing that rational pathway requires understanding the fundamental factors that have precluded the ability to control the underlying pathophysiology of sepsis.

## 1.2 The Challenge of Sepsis

We pose that there are two primary fundamental issues that challenge finding a cure for sepsis: 1) the intrinsic dynamic complexity and redundancy of the immune response and 2) the intractable heterogeneity and variability of the sepsis population. We also believe that these challenges are associated with any disease process that involves the immune response but focus here on the use case of sepsis.

It is crucial to recognize that sepsis arises from the disordered dynamics of an essential beneficial process: the ability of the host to respond to infection or injury. This capability, manifesting as acute inflammation, consists of an initial production and release of pro-inflammatory chemokines and cytokines [5, 6] to kill and contain infectious organisms, the extent of which is concurrently controlled and contained by anti-inflammatory mediators [6, 7]. Given the essential nature of this response, organisms have evolved a robust series of intersecting and redundant pathways to provide this functionality. Sepsis represents the disordered dynamics of these beneficial positive and negative feedback processes where an ostensibly beneficial response leads to excessive collateral damage to otherwise normal tissue (e.g., remote organ failure). However, even in a disordered state the responses seen in this highly connected and redundant system are "locally appropriate," meaning they are functioning as they have evolved to do. As a result, the disease state of acutely disordered acute inflammation seen in critical illness is itself a robust one, and therefore highly resistant to attempts to adjust its trajectory towards a beneficial outcome [8]. This robustness of the pathophysiological state (which can be considered an *attractor* in mathematical terms [8]) has significant consequences given the prevailing approaches to drug-based therapy, which essentially make linear/reductionist approximations of cause and effect and are tested in pre-clinical models designed to reduce the confounding processes and factors that are present in the clinical setting.

This lack of mapping between pre-clinical experimental platforms and the heterogeneity seen in the clinical setting is accentuated by how the issue of dynamic complexity of sepsis widens the gulf between how sepsis patients are identified and potentially classified, and their actual underlying mechanistic biological state. The current means of diagnosing sepsis involves the application of the Sepsis 3 guidelines, a consensus statement that uses only physiological metrics and clinically available laboratory values [9]. In fact, the search for effective biomarkers to classify disease severity in sepsis has been as unfruitful as the search for effective therapeutics; the aforementioned issue of dynamic complexity underlies both these failures. We have computationally demonstrated both the extent of dynamic heterogeneity that is possible in a sepsis population [8] as well as the non-uniqueness of the mapping between a particular system state/configuration and system-level phenotype [10], both of which point to the futility of attempts to create effective predictive biomarker panels for sepsis. To put it another way, sepsis is a *disease state*, not a disease; i.e., sepsis represents dysregulated inflammation as a whole, not along a specific pathway.

The implications of these two insights help define the deficits in current conceptual approaches to curing sepsis and point to what is necessary to effectively treat sepsis: namely that the dynamic state of an individual patient cannot be "fixed" at the onset of the disease, and that the actual type of disorder (and what is required to correct it) will vary over time within a single patient. As a result, any attempt to risk-stratify patients, while potentially useful for prognosis, does not allow for diverging risks/outcomes during the patient's course of disease. For similar reason, the search for genetic "endotypes", which seek to identify risk categories based on -omics profiles, will not address the changing dynamic state that is so readily evident clinically.

In short, what is needed to cure sepsis is the ability to tailor a particular therapy to a specific patient at a specific time during their disease course, in essence "true" precision personalized medicine. We have previously described the formal requirements for "True" Precision Medicine with 4 axioms [11]:

- Axiom 1: Patient A is not the same as Patient B (Personalization)
- Axiom 2: Patient A at Time X is not the same as Patient A at Time Y (Precision)
- Axiom 3: The goal of medicine is to treat, prognosis is not enough (Treatment)
- Axiom 4: Precision medicine should find effective therapies for every patient and not only identify groups of patients that respond to a particular regimen (Inclusiveness)

We have proposed that sepsis can be treated as a complex adaptive control problem [11-13] for which a computational representation of the patient can be used for model-based control. This approach is consistent with the increasingly popular concept of "digital twins," and, we assert, is necessary if sepsis (and all other systemic inflammation syndromes seen in critical illness) are to be cured.

## 1.3 Introduction to Digital Twins and the NASEM Report

The concept of a digital twin (DT) originated in the manufacturing and operations fields in the 1990-2000s as a means of using computer analogs to real world objects or processes to aid in these products' lifecycle management. As defined by one of the originators of the term, a DT comprises of [14]:

- A data structure for the real-world system
- Some process that links data together to form dynamics
- Some link to the real world that feeds back data into the data-propagation/generation process

DTs have found extensive use in industrial applications, both for manufacturing processes and specific engineered systems (e.g., aircraft engines). The key appeal of DTs is that they are "twinned" to a specific real-world object, where the "twinness" between the digital and real-world object is maintained by an ongoing data link between the twins; this allows simulations to be run on the DT to predict the behavior of the real-world twin, anticipating potential failures and proscribing an individualized maintenance plan. However, the intuitive appeal of the DT has led to an explosion of applications using the term, without necessarily being adherent to the requirements of the technology; this has been particularly true in the biomedical arena.

To provide some guidance on research and development of DTs (as applied to multiple domains), the United States National Academies of Science, Engineering and Medicine (NASEM) undertook the creation of an extensive report on this technological approach. In December 2023 NASEM released its report "Foundational Research Gaps and Future Directions for Digital Twins (2023)" after an in-depth evaluation of the potential application of DT to Atmospheric Science, Engineering and Biomedicine [15]. We consider this report (henceforth referred to as the "NASEM Report") to be an authoritative regarding what constitutes a DT that presents a series of unique capabilities and research challenges for the promise of the technology to be eventually met. In the section below we highlight key definitions, findings and recommendations of this report that are pertinent to the development of DT technology that can aid in curing sepsis; we term this aspirational goal a Critical Illness Digital Twin (CIDT).

## 1.4 Highlights of the NASEM Report as pertaining to biomedical DTs such as the CIDT

### 1.4.1 Definition of a Digital Twin

One of the most common questions regarding DTs is "what is the difference between a DT and a computational model"? The NASEM report addresses this with the following text:

"*Finding 2-1: A digital twin is more than just simulation and modeling.*

*Conclusion 2-1: The key elements that comprise a digital twin include (1) modeling and simulation to create a virtual representation of a physical counterpart, and (2) a bidirectional interaction between the virtual and the physical. This bidirectional interaction forms a feedback loop that comprises dynamic data-driven model updating (e.g., sensor fusion, inversion, data assimilation) and optimal decision-making (e.g., control, sensor steering)."* Page 2, NASEM Report

The key point of this definition is that it separates a DT from a predictive computational model, such as an -omics based profile or endotype classifier. While potentially beneficial, those types of predictive models are predicated on a static feature set extracted from a patient at time of the creation of the model, and is not (and often cannot be) updated in an ongoing fashion; therefore these types of predictive models cannot capture disease trajectories or dynamics (and thus do not fulfill Axiom #2 "Precision" of the Axioms of True Precision Medicine), which we have already noted is a key feature of sepsis and DTs.

The other prevalent term that can be distinguished from DTs are the generation and use of virtual populations or in silico cohorts/trials. These approaches [16-19] share many features of DTs: they utilize a shared computational specification (either a mechanism-based dynamic model or a ML/AI correlative model) and reproduce the variation seen in a clinical population. While an incredibly useful technology, particularly in terms of plausibility testing of potential therapies ([16-20]), what is missing in these approaches is the individualization of a particular patient trajectory from within that virtual population, thus failing to fulfill Axiom #1 "Personalization" of the Axioms of True Precision Medicine. The transition from the underlying technology/model specification used to generated a virtual population/perform an in silico trial can be readily accomplished (we will show this in the transition from our work on in silico clinical trials [20] to a proto-DT that proposes to cure sepsis [12, 13]) but it requires the contextualization of the underlying computational model into a use-case that involves employing the ongoing data linkage between the real and virtual twins to explicitly represent an individual patient trajectory.

### 1.4.2 Fit-for -Purpose

The importance of use-case context is the subject of the next emphasis point in the NASEM report. Given the range of possible computational representations of a particular system, the NASEM report emphasizes that DT implementations are heavily influenced by the ostensible use of the DT, e.g., "fit for purpose."

*"Conclusion 3-1: A digital twin should be defined at a level of fidelity and resolution that makes it fit for purpose. Important considerations are the required level of fidelity for prediction of the quantities of interest, the available computational resources, and the acceptable cost. This may lead to the digital twin including high-fidelity, simplified, or surrogate models, as well as a mixture thereof. Furthermore, a digital twin may include the ability to represent and query the virtual models at variable levels of resolution and fidelity depending on the particular task at hand and the available resources (e.g., time, computing, bandwidth, data)."* Page 3, NASEM Report

*"An additional consideration is the complementary role of models and data—a digital twin is distinguished from traditional modeling and simulation in the way that models and data work together to drive decision-making. In cases in which an abundance of data exists and the decisions to be made fall largely within the realm of conditions represented by the data, a data-centric view of a digital twin is appropriate—the data form the core of the digital twin, the numerical model is likely heavily empirical, and analytics and decision-making wrap around this numerical model. In other cases that are data-poor and call on the digital twin to issue predictions in extrapolatory regimes that go well beyond the available data, a model-centric view of a digital twin is appropriate—a mathematical model and its associated numerical model form the core of the digital twin, and data are assimilated through the lens of these models. An important need is to advance hybrid modeling approaches that leverage the synergistic strengths of data-driven and model-driven digital twin formulations."* Page 3, NASEM Report

The implication of these conclusions on the development of biomedical DTs is that such a DT need not be a comprehensive and complete representation of biology. This is fortunate because, as opposed to physical systems for which there are identified natural laws, biological systems exist in a state of perpetual epistemic uncertainty and incompleteness, and *ab initio* $1^{st}$ principles models are not possible for biological systems. Therefore, the design of a biomedical DT should include a level of resolution and abstraction that suits the specific purpose of the DT, and allow for a strategy to deal with the uncertainties regarding the basis of the computational specification of the DT. Thus, while the computational representation is not required to be comprehensive, it must take into account (e.g., through stochastic effects) the epistemic uncertainty necessary to represent the whole system. We have previously outlined a workflow that can provide a guide to how to identify and address the questions/decisions required in order to create a biomedical DT [21]; this flowchart can be seen in Figure 1.

### 1.4.3 Validation and Uncertainty Quantification

The epistemic uncertainty present in the knowledge that goes into a biomedical DT directly impacts the trustworthiness of the DT when it is used, and the NASEM report addresses this crucial point by placing considerable emphasis on the important of Verification, Validation and Uncertainty Quantification (VVUQ) regarding future development, deployment, and maintenance of a DT, as noted in the text below:

*Conclusion 2-2: Digital twins require VVUQ to be a continual process that must adapt to changes in the physical counterpart, digital twin virtual models, data, and the prediction/decision task at hand. A gap exists between the class of problems that has been considered in traditional modeling and simulation settings and the VVUQ problems that will arise for digital twins.* Page 6, NASEM Report.

*Conclusion 2-3: Despite the growing use of artificial intelligence, machine learning, and empirical modeling in engineering and scientific applications, there is a lack of standards in reporting VVUQ as well as a lack of consideration of confidence in modeling outputs.* Page 6, NASEM Report.

*Conclusion 2-4: Methods for ensuring continual VVUQ and monitoring of digital twins are required to establish trustworthiness. It is critical that VVUQ be deeply embedded in the design, creation, and deployment of digital twins. In future digital twin research developments, VVUQ should play a core role and tight integration should be emphasized. Particular areas of research need include continual verification, continual validation, VVUQ in extrapolatory conditions, and scalable algorithms for complex multiscale, multiphysics, and multi-code digital twin software efforts.* Page 6, NASEM Report.

In terms of biomedical DTs, existing practices of software verification are directly applicable to the computational models underlying biomedical DTs. However, we assert that there are specific properties of biological systems that directly limit 1) traditional approaches to validation and UQ and 2) the application of traditional data-centric ML and AI; both of these limitations arise because of observed biological heterogeneity and variability and the inability to formally characterize that heterogeneity because of perpetual data sparsity [22, 23].

Classical validation involves the initial matching and (ideally) subsequent prediction of real-world behavior using a computational model. Additionally, UQ is the assessment of uncertainty in the computational model; for biological models that uncertainty takes many forms:

1. As noted above, there is perpetual epistemic uncertainty in any computational representation. This is true for any knowledge-based mechanism-based model, but also applies to any data centric model (i.e., a model parameterized through the use of ML or AI) because there is perpetual uncertainty regarding whether a particular training data set is sufficiently representative of the entire range of the biological phenomenon being modeled such that the trained data/ML/AI model can adequately generalize (noting that in general ML/AI models do not adequately generalize).
2. The uncertainty also exists in the "real world," where, due to the perpetual sparsity of data relative to the characterizing feature set in biology, the variance in biological data is such that there is no way to know the actual statistical/probability distribution of a particular data type at a particular time. This uncertainty complicates traditional validation because one does not actually know what the computational model is being compared against.
3. Another source of uncertainty is structural uncertainty in a computational model. However, the very nature of biological pathways, which are invariably parallel and redundant, requires representation of this structure if control modalities are to be investigated. Therefore, in contrast to traditional computational modeling where structural uncertainty is to be avoided, in terms of biological systems aimed at identifying control (as would be the case in a CIDT aimed at curing sepsis), structural uncertainty is a necessity in the computational specification.

Because of perpetual epistemic incompleteness in the computational representation and unquantifiable uncertainty in the real world correlate we assert that validation and uncertainty quantification for mechanism-based Medical Digital Twins (MDTs) are entangled. We describe our approach to addressing these challenges in the Methods section.

### 1.4.4. Guiding Sensor Development

The ongoing nature of the data-link between the real and digital twin represents a significant departure from more traditional predictive modeling, particular those that are based on initial conditions that are invariant, such as genetic make-up or demographic/epidemiological properties. Furthermore, patient characterization methods that rely on a tissue sample (such as molecular profiling of tumors) do not allow for subsequent sampling and updating of the biological object being "twinned." Recognizing this current limitation the NASEM Report provided the following conclusion:

*Conclusion 6-1: There is value in digital twins that can optimally design and steer data collection, with the ultimate goal of supporting better decision-making.* Page 64, NASEM Report.

This conclusion is consistent with the use of simulation in many other domains in the physical sciences, where there is model-driven development of experiments or components. This finding also points to the importance of developing multi-disciplinary teams in the development of medical digital twins, and an overall concept of *MDTs as cyberphysical systems that meld software and hardware*.

**1.4.5 The goal of Control**

The melding of software and hardware is also relevant to the next Conclusion of the NASEM Report of control as an integrated part of a DT's fit-for-purpose. This is consistent with our above-mentioned Axiom 3: The goal of medicine is to treat - prognosis is not enough. Again, this should be self-evident, as the intent of the practice of medicine is to aid ill patients to return to a state of "health" (or, more precisely, pre-illness). This goal can be readily described as steering an individual patient's disease trajectory to a state prior to the manifestation of that particular disease, therefore constituting a control task. The dynamic nature of this control task is reflected in the following Finding from the NASEM Report:

*Finding 6-3: Theory and methods are being developed for reinforcement learning and for dynamically adaptive optimization and control algorithms. There is an opportunity to connect these advances more strongly to the development of digital twin methodologies.* Page 66, NASEM Report.

Additionally, the use of an MDT for decision-making, particularly if a closed-loop system is proposed, must have a rigorously established safety profile. Inherent to this evaluation is the ability for practitioners and regulatory bodies to be able to assess *why* a particular control action is suggested by the DT system. Therefore, the process of control discovery using MDTs should be interpretable and explainable (two features notably absent from ML/AI systems).

**1.4.6. Summary of Focused Findings from the NASEM Report**

To summarize what we consider to be key recognitions to be gleaned from the NASEM Report:

- Importance of trajectories: Disease is dynamic, DTs require updating with real-world data. That data must be accessible.
- Personalization: Identification of individual trajectories required identifying applicable configurations of a shared underlying computational DT specification based on updated data feeds. This data must be accessible and allow for refinement of the space of applicable DTs.
- There is no single DT, but rather an ensemble of adequate DT configurations: Biological heterogeneity, data variance and perpetual uncertainty with regards to model structure and data distributions mean that there is no single configuration/parameterization of the underlying DT specification that is sufficient to fit the data. Therefore, the DT is perpetually an ensemble of DTs that suit the data. Ongoing data feeds will refine that ensemble over time; this is akin to the means by which a forecasting cone from a hurricane model is refined with updated data.
- Identification of control is inherent: the goal of medicine is to treat illness and restore a state of health. This is essentially a control problem: how to steer a disordered biological system to one that meets a definition of "health." Therefore, a biomedical DT should incorporate a means of implementing/discovering control (therapies) and have some degree of interpretability such that at least a plausible rationale for the control action is decipherable.

We assert that by and large existing approaches to finding a cure for sepsis (aided by computational tools) do not fit this paradigm and are insufficient for actually finding a cure.

We these factors in mind, we present an initial approximation of a Critical Illness Digital Twin aimed at curing sepsis.

## 2.0 Methods

### 2.1 Determination of Fit-for-Use and Overall Specification

Given the emphasis placed on "fit-for-use" in the NASEM Report (Conclusions 3-1 and 6-1), it is desirable to have guidance as to how to determine the specification and design of an MDT. We have previously developed such a schema, the MDT Design Schema, to formalize the design process of an MDT that provides guidance on the following: regarding "fit-for-purpose", the type of model representation required, specification of methods for VVUQ and the types of data linkages needed for implementation. The MDT Design Schema is particularly suited for the development of what termed are "Drug Development Digital Twins" [21]: MDTs that are intended to discover and optimize control modalities (i.e., therapies) for patients. The flowchart for the MDT Design Schema is seen in Figure 1.

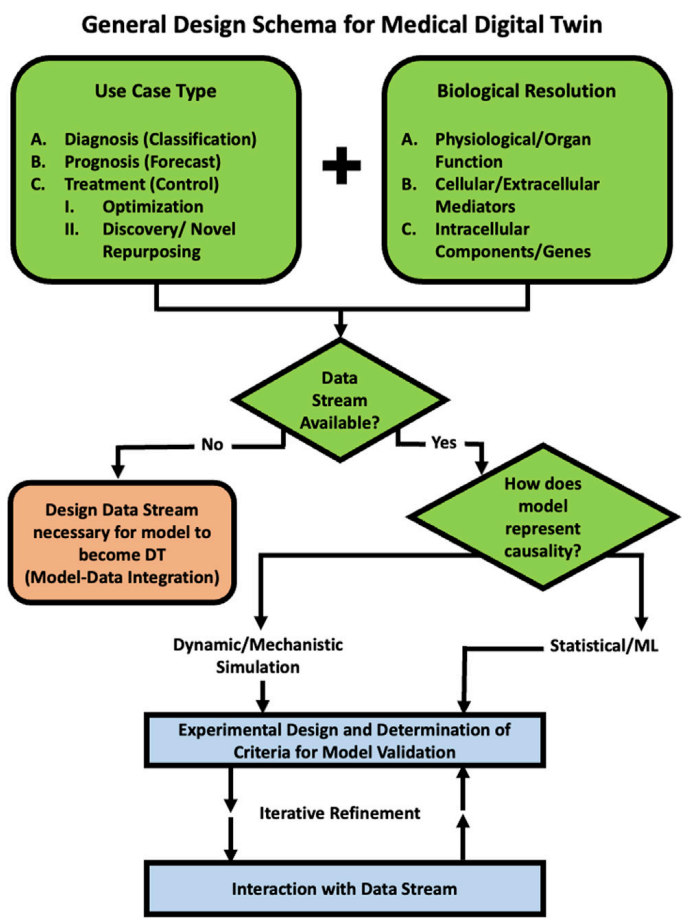

Figure 1: Medical Digital Twin (MDT) Design Schema. This flowchart provides guidance as to the types of decisions that need to be made to develop a MDT that addresses several key findings from the NASEM Report, specifically: "fit-for-purpose", VVUQ, availability and type of ongoing data linkages and possible goal of using the MDT for control. Reconfigured and reproduced from Ref [21] under the Creative Commons License.

As the goal of a Drug Development Digital Twin is to discover putative control strategies to steer a disease trajectory back to a state of health, we apply this schema to the design of a Critical Illness Digital Twin (CIDT) with the explicit goal of being able to cure sepsis through pharmacological manipulation of cellular and molecular mediators (e.g., the application of drugs).

## 2.2 Verification, Validation and Uncertainty Quantitation: New paradigms for MDTs

We recognize that the NASEM Report placed considerable emphasis on the importance of VVUQ for digital twin technology; as noted above VVUQ is the focus of Conclusions 2-2 through 2-4 and Recommendation 2. Specifically, there is a call for new approaches towards what may constitute VVUQ for digital twins based on their specified "fit for purpose." We support this conclusion and have asserted that there are particular properties associated with attempting to computationally represent biomedical systems in a multi-scale fashion, specifically being able to cross cellular-molecular biology to whole patient/clinical phenotypes. We identify two key facts about biological systems that distinguish them from industrial/engineered systems and present intractable challenges to attempts to translate approaches known to be effective in many physical sciences and engineered systems:

- Biological systems are faced with perpetual epistemic uncertainty.

- Biological data, particularly time series data needed to establish system trajectories, will be perpetually sparse with respect to the range of possible data configurations.

We have presented an extensive discussion regarding the implications of these two features of biological systems in Ref [23]; we present a condensed discussion of herein.

We pose an alternative to "uncertainty quantification": rather than uncertainty quantification what is needed is uncertainty conservation, i.e., maintaining maximal information content when mapping between two systems with unquantifiable uncertainty. The uncertainty (perpetual and unquantifiable) in both the specification of the computational model (perpetual epistemic uncertainty) and the probability distributions of the real-world data (data sparsity and biological heterogeneity) are traditionally not addressed in computational models that aim for sufficiency and parsimony in their level of representation/abstraction, and not addressed in the data distribution side by assuming a distribution that has no inherent justification (if an attempt is made to reproduce the variance at all). A partial justification of this traditional approach can be made if the goal of the modeling project is to refine a mechanistic hypothesis or obtain specific knowledge of a particular interaction, but these approaches are fundamentally insufficient if the goal is robust control in a clinically relevant context, as is the case of a DT aimed at curing a disease. For robust control one needs to avoid failure of the proposed control in as a broad a context as possible; we pose that in cases where there are perpetual uncertainties in both the model and the data it is therefore essential to maximize the conservation of information in the how these systems are related to each other, e.g., "communicate" with each other. This insight leads to casting the Validation and Uncertainty description issue for medical DTs, and the Critical Illness DT in particular, in the context of Shannon Information (from Information Theory) and the Maximal Entropy Principle (MEP) from Statistical Mechanics. These are analogous principles where the goal is to account for the uncertainty inherent in systems and their relationship/communication to each other by maintaining the largest set of possible configurations that provide concordance between those systems; i.e., Information as per Information Theory and Entropy as per Statistical Mechanics. Maximizing entropy/information content provides the broadest non-falsifiable set of mechanistic configurations capturing biological and clinical heterogeneity and upon which control can be discovered. Since, however, any putative control policy requires identification of ostensible control points, and any mechanism-based model must have a minimal set of represented components and interactions, we recognize that the discovery of robust control must use a mechanism-based structure that, while acknowledged to be incomplete, needs to be able to be expanded upon to meet the MEP. The task, then, is to design an approach that maximizes the information content given an initial knowledge structure.

Towards this end we have developed a mathematical object called the Model Rule Matrix, or MRM [24]. The MRM is a matrix that consists of columns that list all the entities/molecules/mediators chosen to be included in a simulation model and row that list all the rules representing biological functions/interactions chosen to be in the simulation model. The numerical values of each matrix element denote the strength and direction (inhibition or augmentation) of the contribution of the entity (column) to the functional rule (row). See Figure 2 for a depiction of the components of an MRM. The lack of inclusion of a particular entity in a particular rule is represented by a "0" for the corresponding matrix element in the initial/base MRM; this represents the choices made by the modeler regarding what to include in the computational model, which in turn is informed by what data types are available to link the model to the real-world and provides a set of putative control points.

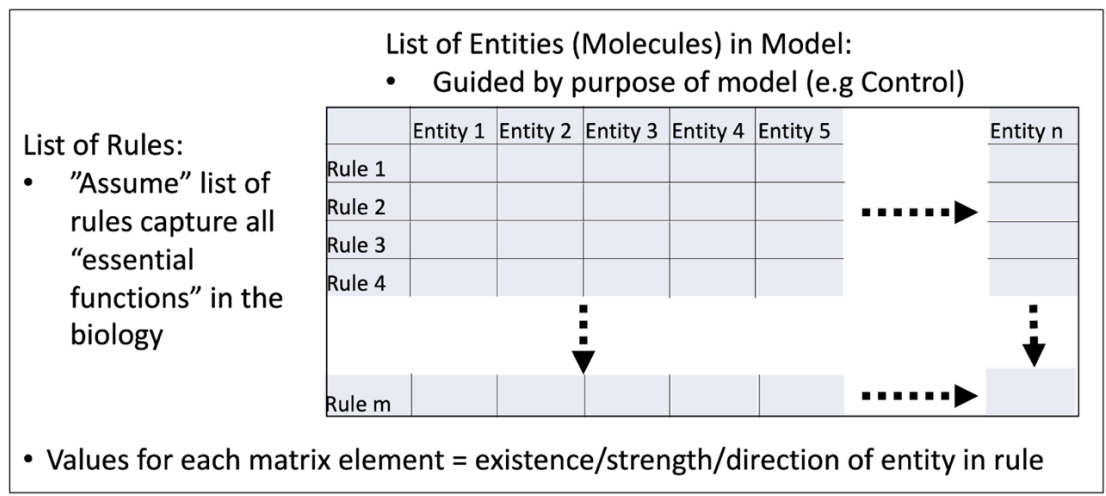

Figure 2: Schematic of Model Rule Matrix (MRM). The MRM is a matrix that depicts the relationship between the entities represented in a mechanism-based computational model (in the columns) and the behavioral rules incorporated in the model (rows). The matrix element values represent a numerical weight and direction (+/-) of the entity in the associated rule. The initial MRM for a given mechanism-based model will only include the explicitly included rules in the model; all the non-represented interactions will have a "0" value for the matrix element. However, these matrix elements represent unknown/unrepresented potential connections/contributions of those entities that might be present in the real-world system; it is these matrix elements that will be uncovered as the MRM is evolved in the below presented machine learning calibration pipeline. Figure reproduced from Ref 25 under the Creative Commons License.

As a result, the base MRM is sparse with most of the possible interactions not explicitly represented in the computational model. However, the selection of what is represented in the model represents a bias with respect to the full biological complexity (though an unavoidable one) and that there are numerous other potential/inevitable contributions from biological components/molecules/pathways/genes that influence the behavior of the explicitly represented model components. Thus the "0" elements in the base MRM represents a representation of a "latent" space of uncharacterized and unspecified interactions that are nonetheless known to be present in some form at some degree. We assert that it is the gap between the explicitly represented structure of a computational model and the recognized additional potential interactions (by whatever degree of connectivity) that contributes to a model to capture the richness (manifested as behavioral heterogeneity) in the real biological system. It is in this fashion that the MRM utilizes the Maximal Entropy Principle: the potential information content of the model is enriched by the representation of connections that are unknown or electively omitted that become necessary to represent the full heterogeneity of a data set. The enriched MRM is then capable of representing a genetically, epigenetically, and functionally diverse cohort of in silico patients able to represent a range of heterogeneous experimental or clinical data.

The process of evolving enriched MRMs involves the application of a ML pipeline that employs both Genetic Algorithms (GAs) [25-28] and Active Learning (AL) [29-33]. to a simulation model constructed such that the coefficients of the rules in the simulation model can be considered strengths of interactions of their associated variables (and is therefore able to be represented by an MRM) and a data set that manifests a large degree of variability. The ML pipeline identifies the set of MRMs (e.g., set of possible additional connectivity configurations) that are able to encompass the range of variability present in the data set. A more detailed description of the ML pipeline can be found at Ref [24, 34]; the workflow is seen in Figure 3:

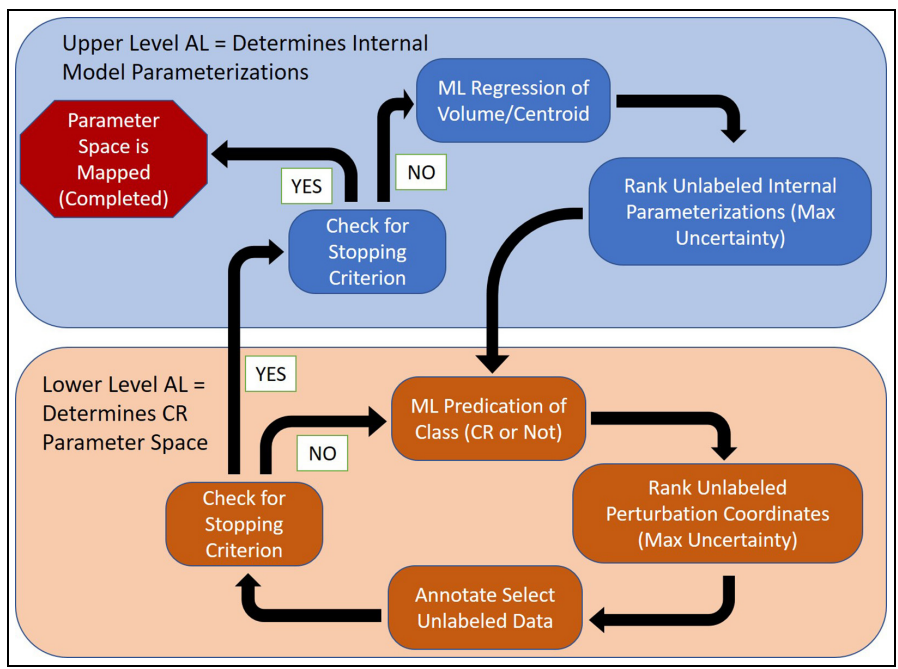

Figure 3: Nested GA/AL MRM Parameter Space Pipeline. This schematic recognizes that there are essentially two distinct classes of parameters when modeling a complex disease such as sepsis or infection. There are parameterizations of the host biology (including potential model structures as reflected in the MRM of the simulation model), and then there are parameterization of the "external" perturbation. In the case of an infection, this refers features of the infection such as the specific properties of the infecting microbe that cannot be quantified or even identified in vivo for a particular organism and the actual amount of the initial inoculum for a particular clinical infection. Therefore, the shown nested structure attempts to identify the clinical relevance (CR) of any single MRM configuration across a space of perturbation features, and then aggregates those CR MRMs to establish the entire non falsifiable MRM space given a particular data set. Figure reproduced from Ref [34] under the Creative Commons License.

The large degree of variability in the data is desirable because the goal is to represent the widest range of biological behavior as possible; having such a target data set expands the descriptive comprehensiveness of an ensemble of MRMs. As opposed to classical parameter fitting, which seeks to find a minimal set (or single) optimal parameter configuration(s), this process does the exact opposite by identifying a very large set of non-falsifiable configurations that has the capability of generalizing beyond the training data set and reducing the risk of brittle, over-fitted models. Time series data is particularly desirable, because the requirement of characterizing of behavioral trajectories further constrains the set of non-falsifiable MRMs (because the knowledge-based form of the rules limits their possible behaviors).

Note that this approach addresses a series of Findings and Conclusions in the NASEM Report concerning fidelity of models, multiscale modeling and model refinement. We consider all these issues related to and inherent in the establishment of novel validity and uncertainty quantification methods such as presented above.

## 2.3 Control Discovery of Complex Systems: Beyond classical control using Deep Reinforcement Learning

While there are numerous classical control approaches to complex dynamical systems, we have previously suggested that for complex biological systems that are represented by agent-based models, such control approaches are limited in their applicability [35]. Instead, we have suggested that Deep Reinforcement Learning (DRL), a form of training ANNs that utilizes time series/outcome data to identify combinations of actions that can alter/change the behavior/trajectories of a system [36], can be applied to simulation models to discover robust control policies to cure sepsis [12, 13]. Our proposed approach is analogous to the successful application of DRL to playing and winning games [37-39]. We term this group of applications of DRL as simulation-based DRL; this describes the use of simulation-generated training data without a pre-defined structural model for the DRL agent. We term this approach *simulation-based DRL*. A key feature of this approach is that since it relies on actions taken on a simulation model there is transparency in the mechanistic state transitions of the simulation system and this allows for a degree of interpretability in the putative control action that is not possible in "standard" data-centric ML and AI systems. This feature has potential significant impact when it comes time to evaluate such a system from a regulatory standpoint on a path to clinical implementation. The details of the process of training a DRL AI can be seen in Ref [13], and are summarized here.

We utilize a Deep Deterministic Policy Gradient (DDPG) [36] to train an AI to cure sepsis. DDPG is a reinforcement learning (RL) algorithm that uses off-policy data and the Bellman equation (Equation 1) to learn a value function, or Q-function, to determine the most valuable action to take given any state of the simulation (obtained by some means of observation, e.g. a sensor in the biomedical case).

$$Q^*(s, a) = \mathop{\mathrm{E}}_{s' \sim P} \left[ r(s, a) + \gamma \max_{a'} Q^*(s', a') \right]$$

Equation 1: The Bellman Equation. Value Q is a function of the current state and action (s, a), and is equal to the reward r from the current state and chosen action (s, a) summed with the discounted value of the next state (discount factor = $\gamma$) and action (s', a') where the next state is sampled from a probability distribution (s' ~ P).

The Q-function is discovered through the execution of numerous simulations of the target system in which the AI agent uses trial and error to optimize the Q-function based on observed rewards from chosen actions. The basis of this process is a discrete version of Q-learning [40], where the next action taken by the AI is chosen from a set of discrete actions with the best action for the current state is the one that returns the highest value from the Q-function. Q-learning is an off-policy algorithm, which means that in the training phase the AI agent is able to choose actions not chosen by the Q-function, allowing the AI agent to explore and potentially discover actions that can lead to a greater reward than an already discovered policy. Q-learning has been demonstrated to be very effective at solving complex control problems in discrete space [36]. The code for the DRL environment (which includes the IIRABM and the DRL training code) can be found at https://github.com/An-Cockrell/DRL_Control.

For the sepsis, our initial application of DRL was intended to provide a proof-of-concept demonstration as to whether sepsis could be theoretically cured at all with this approach [12]. As such, the action and observation spaces for the DDPG were set at clinically unrealistic levels: for observations the state of the IIRABM as reflected by aggregate mediator levels were available every 6 minutes (the length of a single time step of the IIRABM) and the action space allowed every mediator present in the IIRABM (17 in total) to be increased or decreased. The objective function for the DDPG was to reduce total system survival (step-wise reward) and overall system survival (terminal); of note while the total system health/damage was used for the objective function, it was not an observable in executing the DDPG. This initial demonstration proved highly successful, leading to a control policy that reduced a 46% baseline mortality rate to 0.8% in testing over a set of 500 different parameter set/initial condition combinations with baseline mortality rates in the range 1 − 99%. Moreover, no parameter set/initial condition combination exhibited an increase in mortality rate compared to baseline; i.e., no cohort was detrimentally affected by the learned policy. These results suggest that despite being trained on a single patient parameterization, the policy generalizes well, as it is robust to variations in parameter set/initial condition combinations [12].

With these results, we then attempted to more closely approximate a clinically-relevant scenario, with a focus on being able to control sepsis/cytokine storm in the face of a novel pathogen for which no effective anti-microbials existed (ala SARS2-COV) [13]. In this work we wanted to mimic a clinically relevant population and therefore chose parameter values and initial conditions identified from the Nested GA/AL pipeline operating on the IIRABM MRM to provide an overall mortality of ~40%. To make this example more clinically feasible, we set the observation and action space interval to 6 hours, restricted the action space to manipulation of cytokines for which there are currently Food and Drug Administration approved anti-inflammatory biologics, and allowed for the real-time (6 hr. turnaround) assay of the 17 cytokines/mediators included in the IIRABM; this latter mimics what could feasibly done with current state-of-the-art assays [41].

## 3.0 Results

### 3.1 The structure of Critical Illness Digital Twins to cure Sepsis

The application of the requirements of a medical digital twin that fulfills the Axioms of Precision Medicine to our Digital Twin Design Schema leads to the design parameters of a Critical Illness Digital Twin that aims to cure sepsis seen in Figure 4.

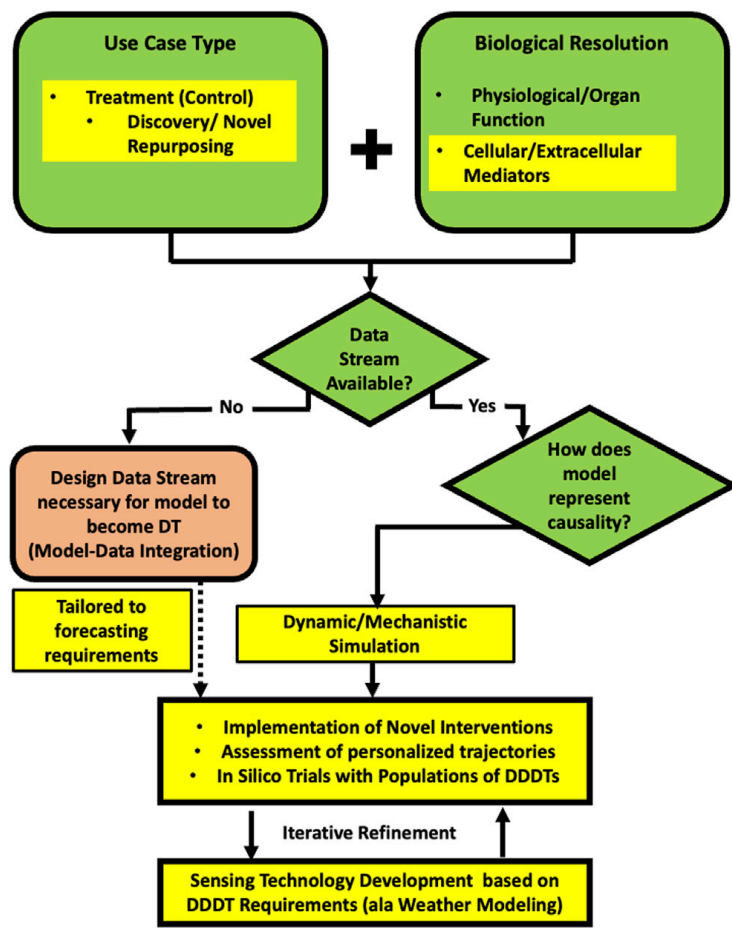

Figure 4. Adaption of the MDT Design Schema to the CIDT, a form of the Drug Development Digital Twin. Yellow boxes represent choices made given the CIDT fit-for-purpose to cure sepsis by discovering new adaptive control strategies based on molecular manipulation (i.e. the application of varying combinations of drugs). Figure is adapted and reproduced from Ref 22 under the Creative Commons License

The design parameters address the following features of the CIDT:

- Use-Case Type/Fit for Purpose: Control discovery, as no current pathophysiologically-targeted therapies exist for sepsis (see NASEM Report Conclusion 3-1 and Finding 6-3).
- Level of Resolution: The ostensible level of control is at the cellular-molecular level, and therefore this determines the resolution needed for the computational specification (see NASEM Report Conclusion 3-1).
- Form of Computational Specification: Since the goal is to test novel interventions, by definition no existing applicable data exists regarding those interventions, and this mandates the use of a mechanism-based simulation model (see NASEM Report Conclusion 3-1).
- Data Requirements: The ongoing data feeds need to link to the representational capability of the underlying computational model, and therefore will be levels of molecular mediators associated with sepsis. We specifically note, however, that this capability does not currently exist, and therefore represents a target for future development (see NASEM Conclusion 6-1)

## 3.2 Computational Specification model for the CIDT: The Innate Immune Response Agent-based Model

Following the guidance schema for developing a MDT we select our previously developed Innate Immune Response Agent-based Model (IIRABM) [8, 20] as the computational specification for the CIDT. The IIRABM has the benefit of being extensively previously validated as a necessary precondition for the ongoing VVUQ required by the NASEM Report Conclusions 2-2 and 2-3. We select the IIRABM for the following reasons:

1. It incorporates an essential set of pro- and anti-inflammatory mediators and associated cell types needed to reproduce the fundamental dynamics of sepsis.
2. Has enough representation of the parallelism present in the response pathways such that trivial solutions are not possible.
3. IT has been used extensively to characterize the complex dynamics of sepsis and the discovery of complex treatment regimens needed to cure sepsis [8, 12, 13, 28].

The IIRABM is a two-dimensional abstract representation of the human endothelial-blood interface with the modeling assumption that the endothelial-blood interface is the initiation site for acute inflammation. The closed nature of the circulatory surface can be represented as a torus, and the two-dimensional surface of the IIRABM therefore represents the sum-total of the capillary beds in the body. The spatial scale of the real-world system is not directly mapped using this scheme. The IIRABM simulates the cellular inflammatory signaling network response to injury/infection and reproduces all the overall clinical trajectories of sepsis [20], and clinically plausible mediator trajectories associated with acute systemic inflammation in response to infection [20, 42, 43]. The IIRABM incorporates multiple cell types and their interactions: endothelial cells, macrophages, neutrophils, TH0, TH1, and TH2 cells as well as their associated precursor immune cells. A schematic of the components and interactions in the IIRABM can be seen in Figure 5.

The IIRABM incorporates stochasticity in a biologically realistic fashion. What this means is that rather than adding an arbitrary noise function to a system of deterministic equations (as would be the case in stochastic differential equations) the IIRABM adds randomness where one would find it in the actual biological system: 1) in the initial spatial distribution of inflammatory cells in tissue, 2) in cell movement not under the influence of chemotaxis and 3) in control points present in branching signaling pathways. This results in stochastic behavior in populations of IIRAB runs that mimic clinical populations [8, 20].

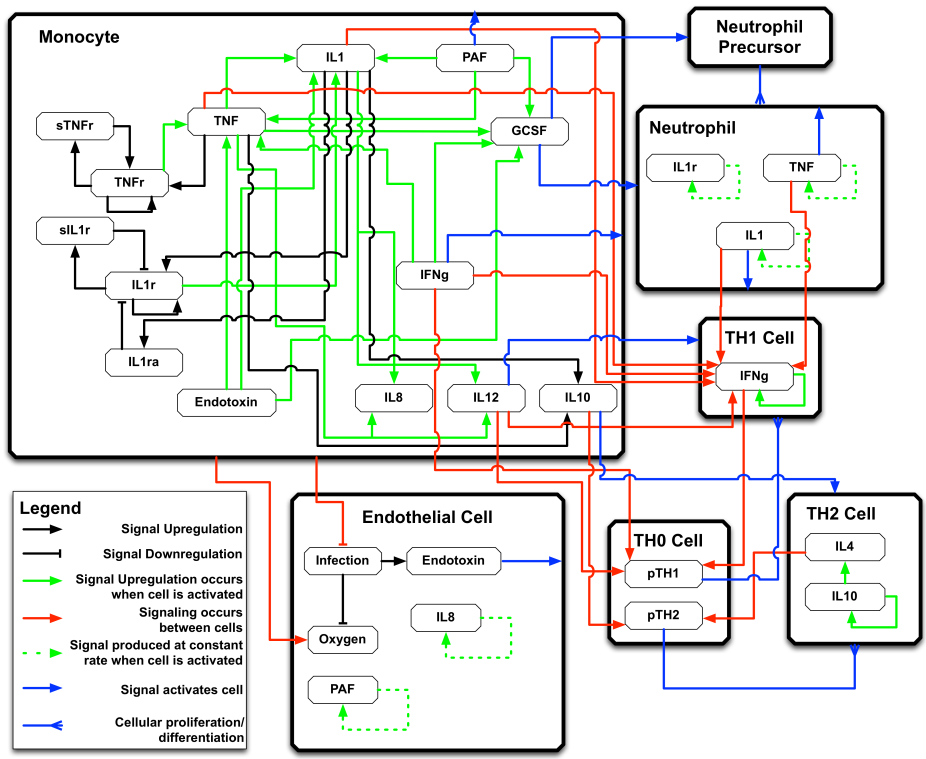

Figure 5. Schematic of cell types, mediator and connections in the Innate Immune Response Agent-based Model (IIRABM). IL1 = Interleukin-1, IL4 = Interleukin-4, IL8 = Interleukin-8, IL10 = Interleukin-10, IL12 = Interleukin-12, TNF = Tumor Necrosis Factor, GCSF = Granulocyte Colony Stimulation Factor, IFNg = Interferon-gamma, TNFr = Tumor necrosis factor receptor, sTNFr = soluble Tumor Necrosis Factor receptor, IL1r = Interleukin-1 receptor, IL1ra = Interleukin-1 receptor antagonist, pTH1 = pro-Type 1 Helper T-cell state, pTH2 = pro-Type 2 Helper T-Cell state.. For full details of the IIRABM see Ref [20]. Figure reprinted from Ref [43] under the Creative Commons License.

### 3.3 "Modified" Validation and Uncertainty Quantification/Conservation

We apply our approach to Uncertainty Conservation by generating a MRM for the base IIRABM and operating on it using our established GA/AL parameter space characterization workflow (see Figures 2 and 3). We present two figures that demonstrate the results of this process. The first is seen in Figure 6, which demonstrates the evolution of the MRM from the base version representing the constructed IIRABM to an ensemble of non-falsifiable MRM configurations; this work is described fully in Ref [24]. Figure 6 shows the difference between the base MRM of the IIRABM (Panel A), which only includes the molecular interactions/rules explicitly programmed into the model, and an example of an evolved MRM (Panel B). The evolved MRM shows a much higher degree of connectivity between the molecular entities, as would be expected in the real biological system (analogous to the interaction map seen in gene network analyses), and the difference is what would be expected given the necessary abstractions made in the construction of a mechanism-based model.

Note that Figure 6B is only one of a series of non-falsifiable (by currently available data) configurations of the IIRABM MRM; the GA/AL pipeline employs a sufficiency criterion where only one of a set of stochastic replicates (n = 100 in this case) need encompass a real-world data point; the rationale for this is the supposition that the maximum value of a mediator in a data set does not represent a bound for human possibility.

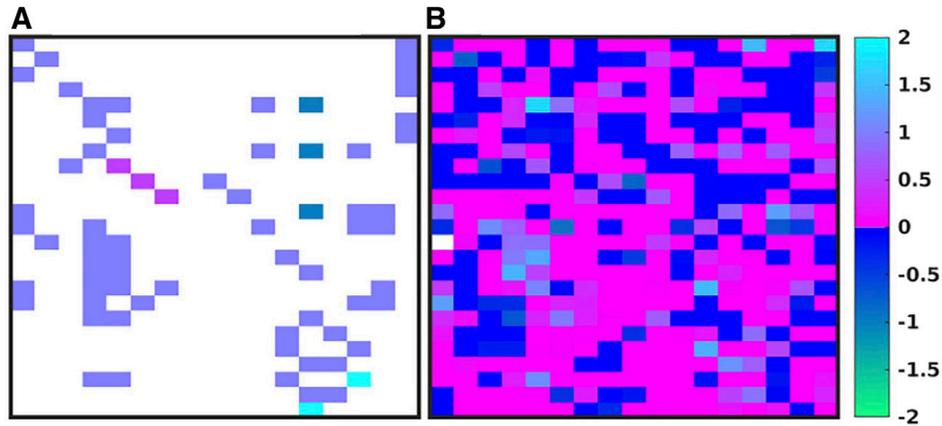

Figure 6: Depictions of the MRM for the IIRABM. The MRM for the IIRABM consists of 17 columns, one for each circulating mediator present in the IIRABM, and 25 rows, one for each molecular rule in the IIRABM. Panel A shows a heatmap of the base IIRABM, which only includes the molecular interactions explicitly represented in the IIRABM. Panel B shows an individual optimized matrix (one of many) representative of the valid ensemble is shown in Panel B. In Panels A and B, the white blocks represent a matrix element with a value of 0 (e.g. no connection); the dark blue to green represents a negative matrix element; the pink to light blue represents a positive matrix element. The optimization process vastly increases the connectivity of the ABM elements as would be present in the actual biological system. Figure reproduced from Ref 25 under the Creative Commons License.

Figure 7 shows what the space of the ensemble of sufficient MRMs looks like.

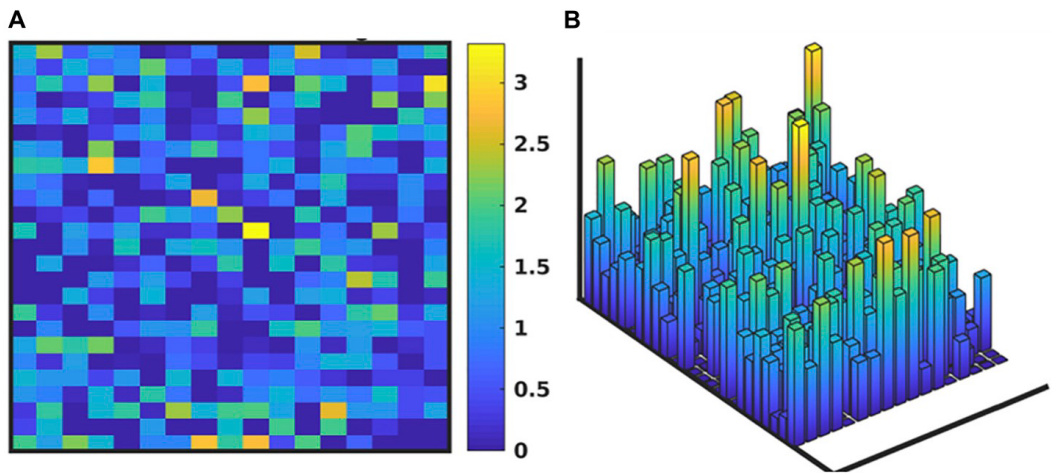

Figure 7: Depiction of space of sufficient ensemble of IIRABM MRMs. Depiction of the range of values of elements the IIRABM MRM for the valid ensemble of MRMs able to produce simulation output non-falsifiable by the clinical data; this is consistent with conservation of maximal information content and entropy as per the MEP. Panel A shows the ranges of the MRM element values as a heatmap, where dark blue is a range of 0 and yellow indicates a range of 3.42, with a maximal range of 4.0. Panel B shows this same data as a 3-dimensional bar graph, where the height of each cell reflects the range of the values for each matrix element. Figure reproduced from Ref 25 under the Creative Commons License.

The process described herein addresses the need to potentially redefine what "validity" and "uncertainty quantification" mean in the context of a biomedical DT developed with the goal of discovering robust control (as per NASEM Report Conclusions 2-2, 2-3 and 2-4). We note again that this task needs to reconcile two systems with unquantifiable uncertainty; we assert in this type of case validity is established in terms of uncertainty conservation. Because the goal of the DT is to provide a means for robust control, no one MRM configuration can be said to be invalid (based on this data set), therefore remains a possible configuration corresponding an individual patient and thus needs to be accounted for in the process of robust discovery. We wish to note that while the set of sufficient MRMs is very, very large (essentially the inverse of fitting parameters to a mean, as is traditionally done), it is not infinite, and can be seen in Figure 8 to have a discernable structure.

We assert that the greatly expanded trajectory space provided by identification of sufficient MRMs more comprehensively captures the set of possible behaviors present in the clinical population (again, due to the inherent sparsity of clinical sampling). Our second example is fully reported in and demonstrates this extension of potential patient trajectory space by applying the GA/AL pipeline to the IIRABM MRM and a clinical data set from the Uniform Services University/Walter Reed National Medical Military Center. This data set consists of 199 trauma patients, 92 of which developed Acute Respiratory Distress Syndrome (ARDS) at some point during the course of their hospitalization, and 107 controls that did not develop ARDS. This data included time series data of plasma mediators: Interleukin-1-beta (IL-1b), Interleukin-1 receptor antagonist (IL-1ra), Interleukin-6 (IL-6), Interleukin-4 (IL-4), Interleukin-8 (IL-8), Interleukin-10 (IL-10), Granulocyte Colony Stimulating Factor (GCSF), Interferon-gamma (IFNγ), and Tumor Necrosis Factor-alpha (TNFα), and clinical data needed to determine overall pulmonary health state as reflected by the lung Sequential Organ Failure Assessment (SOFA) score: partial pressure of oxygen, complete information regarding respiratory support, and blood oxygen saturation. Figure 9 (reprinted from Ref [44]), demonstrates how the mediator trajectory space is enhanced using the ensemble of IIRABM MRMs.

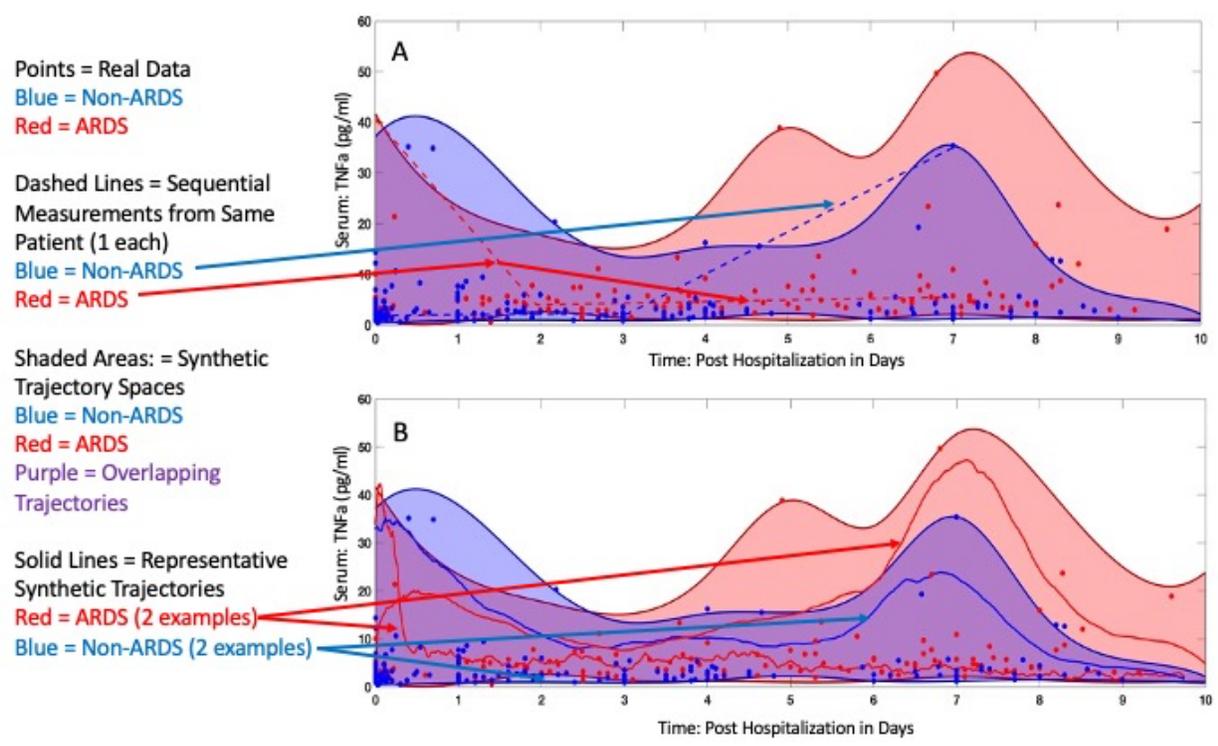

Figure 8: Expansion of clinical mediator time series data with simulated trajectories from ensemble of IIRABM MRMs. In Panel A, clinically collected blood-serum cytokine measurements for TNFα are seen as distinct points, where red points representing patients that developed ARDS and blue points representing patients that did not. The dotted lines indicate representative clinical trajectories of individual patients; note the sparsity of time points for an individual patient's time series. The shading indicates the boundaries of the model trajectory space for the parameterizations that generate ARDS (red) or not (blue). Note significant overlap between the two spaces given the overlap of data points. However, the key

point is that differential parameterizations are able to identify clear regions that are unique to each group. In Panel B, the three solid lines (two red lines eventually developing ARDS and the blue line not) show actual simulated trajectories of TNFα blood-serum concentrations; note here the continuous nature of the trajectories, which would more accurately reflect the actual underlying biological behavior. Reconfigured from Ref 36 under the Creative Commons License.

While Figure 8 demonstrates how the IIRABM MRM ensemble is able to fill out the trajectory space for TNFα, the Nested GA/AL pipeline does this across all the measured cytokines that are represented in the IIRABM. Figure 10 Panels A and B show this same process for GCSF and IL-10, respectively. We are also able to correlate these synthetic CIDT-generated time series to organ-level state in a multi-scale fashion; this is seen in Figure 10 Panel C, where the different groups of patients (ARDS versus non-ARDS) show development of ARDS as trajectories of lung SOFA scores corresponding to mediator trajectories (figure modified from that published in Ref [45].

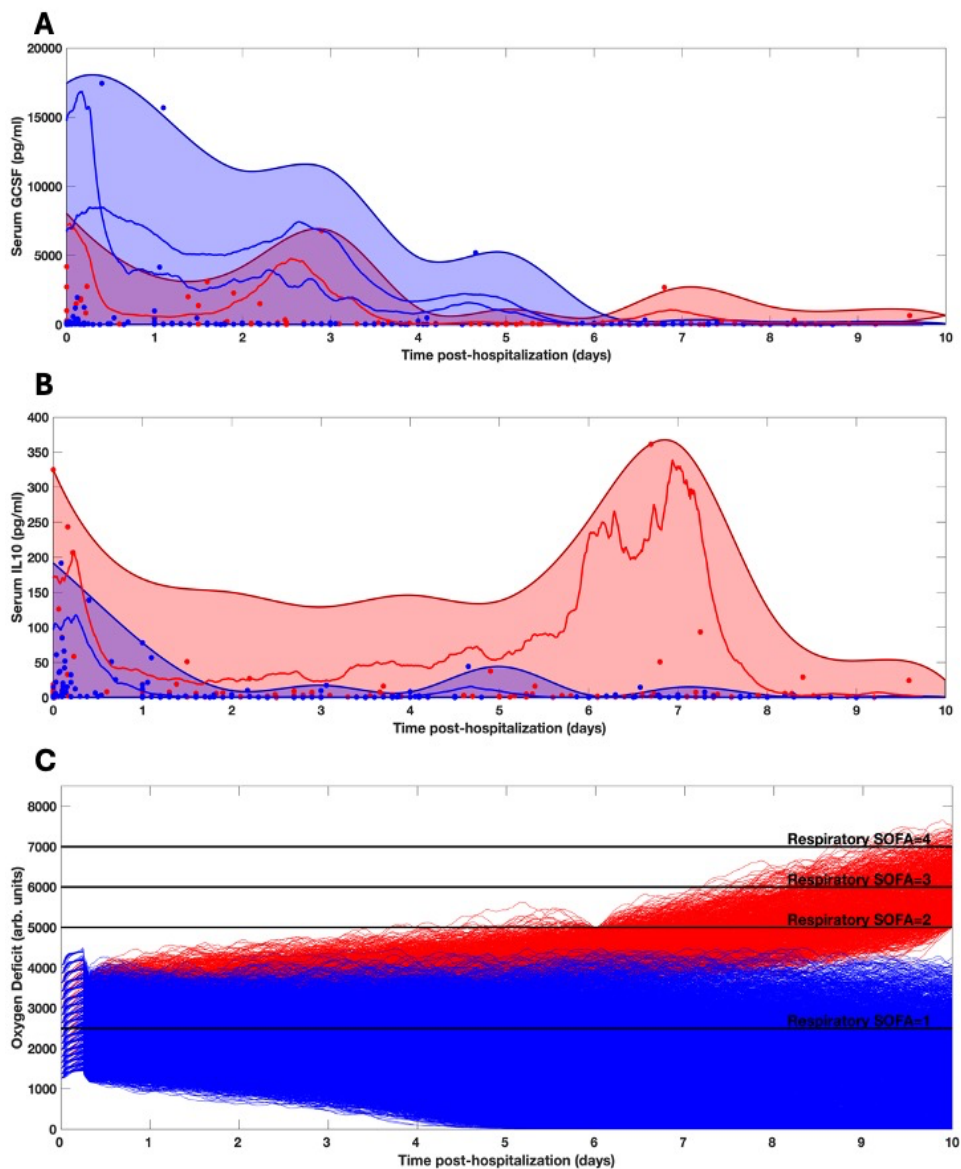

Figure 9: Correspondence Between Mediator Time Series and Simulated Clinical State. Distinct points are the clinically collected data: red points representing patients that developed ARDS and blue points representing patients that did not. The shading indicates the boundaries of the model behavior space for the parameterizations that generate ARDS (red) or not (blue). Panel A: Simulated trajectories for GCSF. Panel B: Simulated Trajectories for IL10. Panel C shows lung damage trajectories of simulated patients and their corresponding respiratory SOFA scores as they develop ARDS (increasing oxygen deficit) or progress towards successful healing of the lung injury. Modified and reproduced from Ref 35 under the Creative Commons License.

A key capability provided by the characterization of the MRM with the Nested GA/AL pipeline is that it reconciles population-level coverage (by the expanded representation of system possibility space provided by the sufficiency criteria in the Nested GA/AL pipeline) while providing a means to replicate and forecast individual trajectories (by identifying specific ensembles of MRMs that encompass a given data point). This latter capability is what sets this approach apart from other methods of generating virtual/in silico populations as means of representing digital twins: it accounts for the uncertainty present in an individual trajectory by providing a rolling forecasting cone that incorporates biological aleatory and epistemic stochasticity and requires an ongoing data linkage with the real world. We note that the resulting simulated data can be used a synthetic training data for machine learning and artificial intelligence methods for both more traditional ML/AI predictive algorithms [44, 45] that focus on generating just such a rolling patient forecast, however we have added an additional novel application of this expansive synthetic data specifically related to the fit-for-purpose of the CIDT to cure sepsis: the application of simulation-based deep reinforcement learning (DRL) for robust policy discovery.

## 3.4 Complex Control Discovery: A DLR-trained AI Control Policy to Cure Sepsis

As noted above we show a more clinically relevant example of our proposed CIDT to cure sepsis using existing anti-cytokine agents to cure sepsis arising from a microbial infection for which no effective anti-microbials exist, such as would be the case with an infection from a multi-drug resistant microbe or a novel pathogen in an emerging pandemic; the extensive details of this work can be seen in [13] and are summarized here.

In a simulated patient course the AI agent initiates intervention 12 hours post application of initial infection; this reflects an approximation of how long it would take for a patient to be ill enough to be recognized as being infected, and the time course simulated = 21 days. Training is performed on a parameterization with 39% Mortality; we will evaluate the generalizability of the learned treatment policy by testing it on a series of additional parameterizations. The actions taken by the AI agent can either be augmentation or inhibition of six mediators present in the IIRABM for which there are existing FDA-approved pharmacological agents: Tumor Necrosis Factor-alpha (TNF$\alpha$), Interleukin-1 (IL-1), Interleukin-2 (IL-2), Interleukin-4 (IL-4), Interleukin-8 (IL-8), Interleukin-12 (IL-12) and Interferon-gamma (IFN$\gamma$). As this study is a proof-of-concept for potential clinical plausibility, we choose a hypothetical yet clinically plausible time frame in which a potential blood mediator assay would be run and used to inform the administration of a drug/set of drugs of 6 hours (see Ref [41] for potentially applicable technology). As a simplifying approximation of clinical pharmacological effect the duration of the effect of each intervention was simulated to last for 6 hours.

The results of the DRL on the CIDT with these clinically feasible constraints on the observation and action space provided a trained AI with a policy that reduced a baseline mortality rate of 39% (61% Recovery) to 10% (90% Recovery). Notably, this control policy completely eradicated the initial infection in all cases, demonstrating that controlled augmentation of the immune system could clear an infection with the aid of antimicrobials. The AI learned policy involved manipulating each of the 6 targeted mediators in some fashion (either augmentation or inhibition) every six hours in a variable fashion; the variation in the applied controls demonstrates the variability seen in individual patient trajectories and therefore the need to have constant updating via a data link to evaluate the system's response to a particular applied control.

To aid in visualizing how the implementation of a control policy addresses variability/heterogeneity in cytokine dynamics seen between individuals, the following graphs show the specific control trajectories for three specific simulated individuals who were successfully treated (Figure 11).

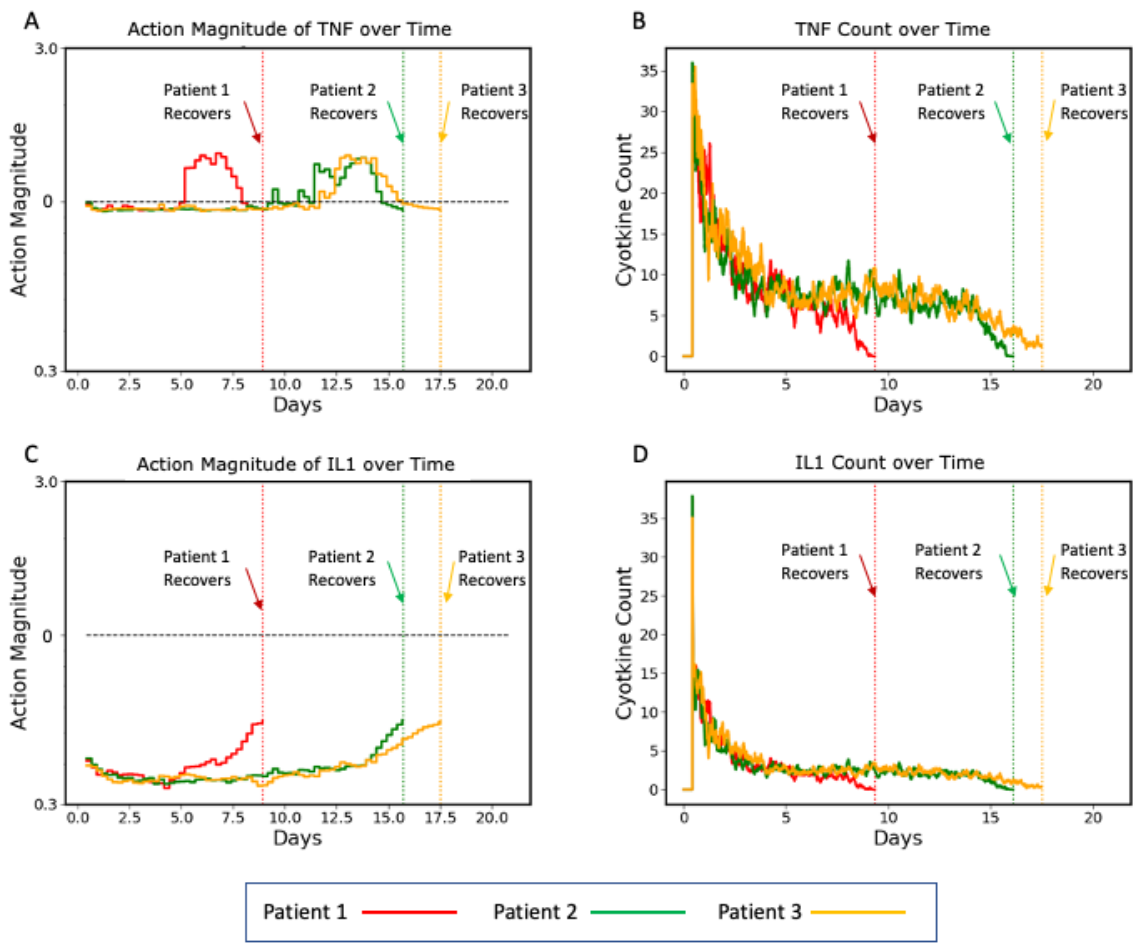

Figure 10 – Plots of three representative individual controlled simulated patients, all of which recover, showing control actions (Panels A and C) and targeted mediators being controlled (Panels B and D). Panels A and B show the control actions on TNF and the resulting trajectories of TNF, respectively. Panels C and D show the control actions on IL1 and the resulting trajectories of IL1. The different dynamics between the three patients can be seen in the variation between their individual control actions and targeted mediators. Note that each of these simulated patients recovers at a different time point (vertical dashed lines), at which time their respective simulations terminate.

There is an added benefit to this approach as the control is discovered on a mechanistic simulation model and therefore the control actions are interpretable. The full discussion on how the DRL trained AI control policy can be interpreted in this example is beyond the scope of this paper but can be seen in the original reference publication [13].

A frequently voiced concern regarding the use of DRL methods for control discovery is that they are not robust, and too dependent upon how well the simulation model maps to the real world. We acknowledge this challenge and pose that our approach that uses the non-falsifiable MRM space and consequent conservation of information in identifying that space can potentially mitigate this. While we do not currently have the means to confirm the adequacy of virtual-to-real world mapping in terms of the discovered control policy (see Discussion for Future Directions) we do demonstrate that the discovered control policy is robust across a wide breadth of parameterizations of the CIDT. As a demonstration, in Ref [13] we show that the DRL discovered control policy is able to manage a range of parameterizations reflecting a wide range of values of microbial properties (i.e. Invasiveness = the invasiveness of the organism and Toxigenesis = the cellular toxicity of the organism), host factors (Host Resilience = rate at which damaged cells recover) and severity of initial infection (Initial Injury); note that these correspond to the "lower" nested level of the GA/AL pipeline seen in Figure 4. This set of 4 Test conditions are as follows:

- Test 1: Host Resilience = 0.1: Invasiveness = 1:Toxigenesis = 3: Initial Injury = 20. Notation Label (0.1, 1, 3, 20). Test 1 Parameters generated a Baseline Recovered Rate = 25%. This parameterization represents a group with higher health resilience (corresponding to better baseline health status), but exposed to a microbe that more rapidly kills infected cells.
- Test 2: Host Resilience = 0.12: Invasiveness = 1:Toxigenesis = 1: Initial Injury = 32. Notation Label (0.12, 1, 1, 32). Test 2 Parameters generated a Baseline Recovered Rate = 16%. This parameterization represents a group with even higher baseline health, but is exposed to a much larger initial inoculum (demonstrating dose-dependent disease severity).
- Test 3: Host Resilience = 0.08: Invasiveness = 2:Toxigenesis = 1: Initial Injury = 23. Notation Label (0.08, 2, 1, 23). Test 3 Parameters generated a Baseline Recovered Rate = 19%. This parameterization represents a group with the same baseline health status as the training set, but was exposed to a microbe that more readily infects surrounding cells.
- Test 4: Host Resilience = 0.12: Invasiveness = 2:Toxigenesis = 1: Initial Injury = 28. Notation Label (0.12, 2, 1, 28). Test 4 Parameters generated a Baseline Recovered Rate = 37%. This parameterization represents a group with higher baseline health, but was exposed to a microbe that more readily infects surrounding cells.

The results of applying the previously trained DRL-policy is shown in Table 1, demonstrating generalizability across these Test conditions with improvements in Recovered Rate ranging from +33% to +56%.

| Parameterization | Uncontrolled Recovered Rate | Controlled Recovered Rate | Improvement |
| --- | --- | --- | --- |
| Test 1: (0.1,1,3,20) | 25% | 81% | 56% |
| Test 2: (0.12,1,1,32) | 16% | 56% | 40% |
| Test 3: (0.08,2,1,23) | 19% | 52% | 33% |
| Test 4: (0.12,2,1,28) | 37% | 83% | 46% |

Table 1 - Therapeutic Model Generalizability: The baseline Recovery Rates and controlled Recovery Rates for IIRABM/CIDT parameterizations upon which the DRL algorithm was not trained. See Text for description of Test Sets 1-4. Table reproduced from Ref [13] under the Creative Commons License.

## 4.0 Discussion

### 4.1 Relationship between Methodology of CIDT to the findings of the NASEM Report

In summary, the approach we propose for developing a CIDT to cure sepsis addresses numerous findings from the NASEM Report. Specifically:

- Meets the NASEM definition of a Digital Twin, where design of the CIDT is such that the dynamic computational model underpinning the CIDT is linked to the real-world twin through an ongoing bidirectional data feed.

- Addresses the need to specify "fit-for-purpose" of the Digital Twin, which is used to guide defining resolution level of the CIDT.
- Addresses the need to develop a comprehensive approach to validation and uncertainty quantification for Digital Twins that recognizes the unique challenges of reconciling unquantifiable uncertainty in both the virtual and real-world twin, and provides a solution governed by Information Theory and Statistical Mechanics manifesting in the MRM.
- Explicitly addresses the issue of control as an ostensible purpose of the Digital Twin, and provides an approach to use simulation-based DRL for control discovery in a system that challenges traditional control theoretic approaches.
- Explicitly utilizes the concept a data linkage between the real- and virtual world to address the "rolling forecast cone" and execution of adaptive control.
- Explicitly defines future directions for sensor development needed to facilitate the ongoing data-linkage requirement.

We propose that the presented components of a CIDT are with the findings and conclusions for DT design, development and deployment put forth in the NASEM Report and represents a pathway towards actually curing sepsis. However, recognize that it currently represents a theoretical solution, and that there are several steps necessary to prove its efficacy and potential gaps that may need to be addressed and represent future directions for investigation and development; some of these are listed in the sections below.

## 4.2 What is needed next for the CIDT to be deployed?

### 4.2.1 Moving from theoretical to real-world efficacy: Real-world testing of the CIDT-derived AI Controller

A recognized potential limitation of this approach is the potential that the underlying model specification of the CIDT is insufficiently representative of the real-world. While we address this issue directly in our approach to VUQ using the MRM, we also recognize that testing a particular knowledge structure under a series of perturbations is critical to refining that knowledge structure. However, given the complexity of the mechanisms being represented, and the complex variable behavior present in the system, we pose that this refinement can only effectively be done in the context of deployment of the complete CIDT-control discovered system. Doing so will require an in vivo testing environment that involves an experimental model of sepsis of sufficient complexity that mirrors clinical situation. The reasons for this are:

- The physiological state space of the experimental model must incorporate current state-of-the-art therapeutic interventions. This is because we assert that clinical sepsis arises primarily in conditions where the individual would have otherwise already died without such support and manifests in cellular-molecular configurations that would otherwise never be seen [46, 47].
- The experimental model must be able to reproduce the heterogeneity and variability of the clinical population. The key point here is that for the MRM and our DRL training strategy to be effective it needs to be exposed to a wide as possible representation of the "possibility space" of the particular organism. This is contrary to the traditional experimental design approach that intends to limit variability, but in this case capturing as much of that heterogeneity as possible is essential [22].

Figure 11 presents a proposed schema for an *in vivo* testing platform integrated with the CIDT-driven DRL training of an AI controller to cure sepsis. There is an inherently iterative process between ongoing refinements of the CIDT and subsequent DRL training based on feedback from the results of the *in vivo* implementation of those controls. Figure 12 depicts a first approximation of this iterative cycle, with sensors consisting of existing multiplexed plasma mediator levels and a multi-channel infusion system that would deliver the specified mediators/inhibitors based on the AI control policy. While theoretically implementable as a closed-loop system, for potential regulatory and safety assessment purposes a human could be readily inserted into this loop to execute the proposed policy actions.

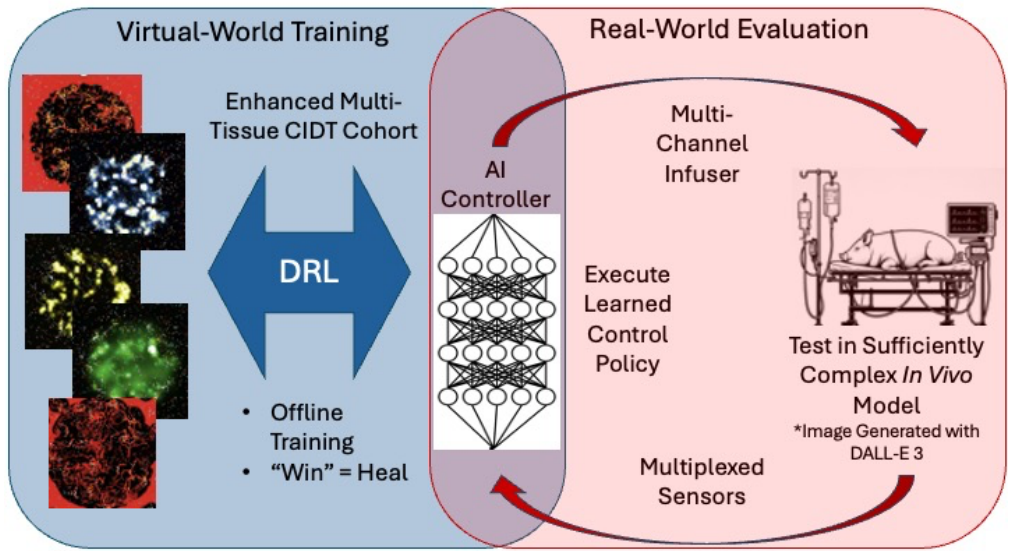

Figure 11: Schematic for *in vivo* test platform for CIDT-trained AI Controller. The processes on the blue shaded left of the figure depict the development of the computational aspect of the CIDT and the simulation-based DRL that trains the AI controller. The red shaded right of the figure depicts the proposed desired *in vivo* testing system of the AI controller.

The iterative nature of this schema mirrors the classical iteration between computational simulations and real-world experiments, with the expectation that performance improvements would be an expected consequence of the deployment of this system, akin to the practice of a "beta release" using a real-world implementation (in this case with a sufficiently complex *in vivo* intensive care model) to refine the system.

### 4.2.2 Addressing potential discrepancy between plasma and tissue molecular pathophysiology: do our sensors and data feed reflect the targets of control?

Inherent to the NASEM Report Definition of digital twins is the presence of an ongoing data link/feed between the real world and virtual objects. In the case of the proposed CIDT the source of information from the real world is the blood, which is readily accessible. However, it is recognized the circulating blood may not reflect the active biology that leads to organ dysfunction; what is present in the blood represents the "spillover" from these tissue-level processes. Notwithstanding the fact that the circulation represents the means by which individual organs communicate with and influence each other, ideally, new sensing technologies (as suggested by the NASEM Report Conclusion 6-1) able to provide repeated ongoing information on cellular actions and outputs in specific organs need to be developed and refined [48]. Some possibilities include high-resolution Magnetic Resolution Imaging (MRI) with (ultrasmall) superparamagentic iron oxide particles (SPIO and USPIO) that can distinguish between types and distribution of inflammatory cells [48, 49], the use of fluorescent dyes that be optically visualized *in vivo* [50], two-photon excitation lasers [51], tissue-localizable near-infrared-fluorescent single-walled carbon nanotubes [52] and aptamer technology [53]. The development and testing of such technologies could ideally be integrated into the experimental system depicted in Figure 12, with concurrent partitioning of the computational model underlying the CIDT (akin to a modular structure proposed in 2008 [54]).

### 4.2.3 Patient Sub-stratification: can we utilize disease classification methods to reduce the candidate trajectories?

There have been extensive attempts to identify means of patient sub-stratification in sepsis [55-59]. We have demonstrated the futility of these attempts due to the non-uniqueness of phenotype and microstate in individual patient trajectories [8, 10], findings that prompted our recognition of DRL as a means of addressing this issue. However, even though relying on cohort sub-stratification cannot guide individualized therapeutic decisions, we do see that there is a potential benefit of clustering patient response tendencies to reduce computational overhead and providing a more directed identification of suitable ensembles that correspond to an individual patient.

### 4.3 Potential issues that are mitigated in the CIDT

#### 4.3.1 What about genetic predisposition?

Emphasis is on function responsiveness of the system, and this is manifest by the updating of the DT through its virtual-physical data link. Given insight from MRM, it is impossible to prespecify/classify/categorize immune responsiveness level given the non-uniqueness of biological trajectories. Therefore, a DT concept with ongoing system updating is required to predict and treat sepsis. The functional responsiveness cannot be parsed into the contributing components (i.e. genetic predisposition, epigenetic status, comorbidities, time course of disease) and therefore there is a mandatory need to capture individualized/individualizable dynamics/trajectories with the DT where there is invariable crossover in the mediator-state trajectories that can arise from different MRM configurations (and therefore reflect different genetic/epigenetic/comorbid predispositions).

#### 4.3.2 What about ethical issues?

When discussing MDTs there is invariably concern regarding ethical issues related to patient privacy, data/DT ownership and equity in the representation of the DT. In fact, the NASEM Report emphasizes these issues in its findings and conclusions:

*Conclusion 6-3: While the capture of enough contextual data detail in the metadata is critical for ensuring appropriate inference and interoperability, the inclusion of increasing details may pose emerging privacy and security risks. This aggregation of potentially sensitive and personalized data and models is particularly challenging for digital twins. A digital twin of a human or component of a human is inherently identifiable and poses questions around privacy and ownership as well as rights to access.* Page 70, NASEM Report.

*Conclusion 6-4: Models may yield discriminatory results from biases of the training data sets or introduced biases from those developing the models. The human-digital twin interaction may result in increased or decreased bias in the decisions that are made.* Page 70, NASEM Report.

We propose that our approach to the CIDT mitigates many of these concerns, specifically due to:
- Our recognition of the lack of a one-to-one mapping between clinical state and underlying cellular-molecular microstate.
- The overlapping variability of microstates, particularly in time series.
- Our reliance on aggregate microstate time series trajectories as the means by which an individual patient is "twinned."

The concept of shared model structure but different parameterizations allows for representation across the range of human responsiveness. Since the personalization process does not rely on traditional statistical significance (instead using sufficiency as determined by the ability of at least one member of a set of stochastic replicates to encompass a data point, as a criterion for identifying a sufficient MRM space) the effects of systematic under-representation are not present.

In a related fashion, traditional/current concerns regarding patient privacy/data ownership are not as relevant as the CIDT is focused on functional responsiveness of a shared underlying biology, recognizing that there is no supposition of uniqueness regarding a particular functional state (in fact, it is expected that a particular functional state could be seen across a host of different patient factors normally considered particularly sensitive). Therefore, the particular configuration(s) of the CIDT that are personalized to a particular patient at a particular time are no different than their laboratory values or monitor outputs; these are PHI but only in the context of all the other factors that determine PHI (i.e., personal information, dates of admission, etc.).

**4.3.3: What about computational overhead and efficiency?**

The NASEM Report notes potential concerns regarding the ability for DTs to support decision making in "real-time" ("Digital Twin Demands for Real-Time Decision-Making" Page 65) and as noted in:

*"Finding 6-1: There is a need for digital twins to support complex trade-offs of risk, performance, cost, and computation time in decision-making."* Page 64, NASEM Report.

This is not a concern for the CIDT as currently designed, as the heavy computational cost is in off-line, both in terms of identifying the sufficient MRM space through the Nested GA-AL pipeline and the DRL training of the controller AI. While there may be limits in the turn-around time for the potential multiplexed plasma mediator sensors, current technology is able to deliver the ~ 6 hrs timeframe that was demonstrated to be effective in the example presented above and in Ref [13].

**4.4 Moving beyond sepsis:**

The focus of the CIDT is critical illness involving acute inflammation, but we view inflammation as a unifying biological and pathophysiological process. In fact, we assert that nearly every significant acquired disease process can be described as a disordered state of inflammation and immunity. Figure 12 is a depiction of the type of disordered inflammation/immunity can be linked to a wide range of disease types.

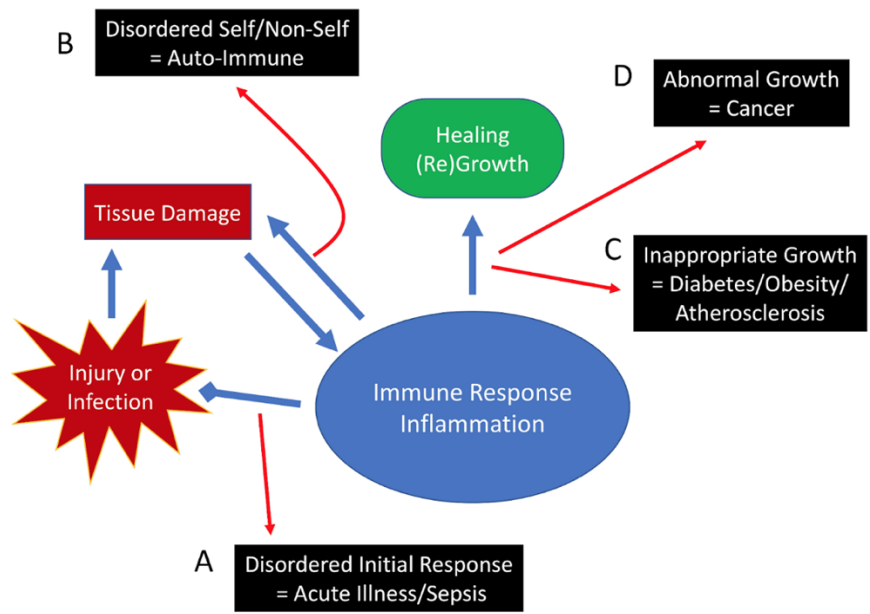

Figure 12: Immune/Inflammatory functions and their relationship to various classes of diseases: A) Acute illness or sepsis occurs with either the initial immune response is unable to contain the infection or, while attempting to do so, causes enough tissue damage such that organs fail (i.e. cytokine storm/cytokine release syndrome). B) Autoimmune diseases in which failure in non-self recognition or negative feedback control of proinflammation leads to persistent inflammation and long-term tissue damage. C) Diseases in which perpetuating ongoing inflammation leads to chronic inflammation and inappropriate growth (as in obesity or adult-onset diabetes. D) In cancers, where chronic inflammation can contribute to initiation of oncogenesis, tumor growth that avoids immune clearance, immune activity that promotes metastasis and an ongoing pro-mutational environment that can lead to resistance. Immune pathophysiological processes range in time scale from hours for acute illness and sepsis to years and decades in autoimmune diseases and cancer. We pose that nearly every disease process and its potential resolution involves inflammation and immunity.

A full discussion regarding this potentially controversial assertion is beyond the scope of this paper, but with respect to the methods and design schema presented herein for the CIDT, given how these approaches fundamentally address the issues raised in the NASEM Report the schema presented herein should be readily suitable to any of the disease types listed in Figure 12.

### 4.5 A sustainable and updateable digital twin ecosystem

Chapter 7 of the NASEM Report is entitled "Toward Scalable and Sustainable Digital Twins" (Page 72, NASEM Report). This chapter acknowledges that there will necessarily be a succession of DTs as technologies (both on the modeling and sensor/actuator sides) improve, there is better understanding of the underlying biology in a DT and as a DT's specific use-cases may evolve. Ideally, the design, development and deployment principles involved in the first-generation DTs will be grounded at a fundamental enough level such that the evolution of these systems represent in-kind refinement as opposed to comprehensive re-conceptualizing and re-engineering of the DT. We believe that the guidance provided by the NASEM Report does address fundamental issues regarding the design, development and deployment of a MDT, and that these fundamentals can be embedded into a larger DT ecosystem that can add an additional layer of potential robust development. A list of these components would include:

Automated knowledge to model generation as new knowledge becomes available. New knowledge will continue to be generated, and this new information needs to be incorporated into new MDTs, while also having backward compatibility with prior capabilities of earlier DTs. This is a knowledge acquisition and integration task that would ideally be integrated with a means of accelerating model construction and evaluation. Many researchers are working in this area, and our group has proposed a potential workflow to facilitate the automated conversion of scientific knowledge into scientific simulation models called the Machine Assisted Generation, Calibration, and Comparison (MAGCC) Framework [60]. A schematic of this framework can be seen in Figure 13.

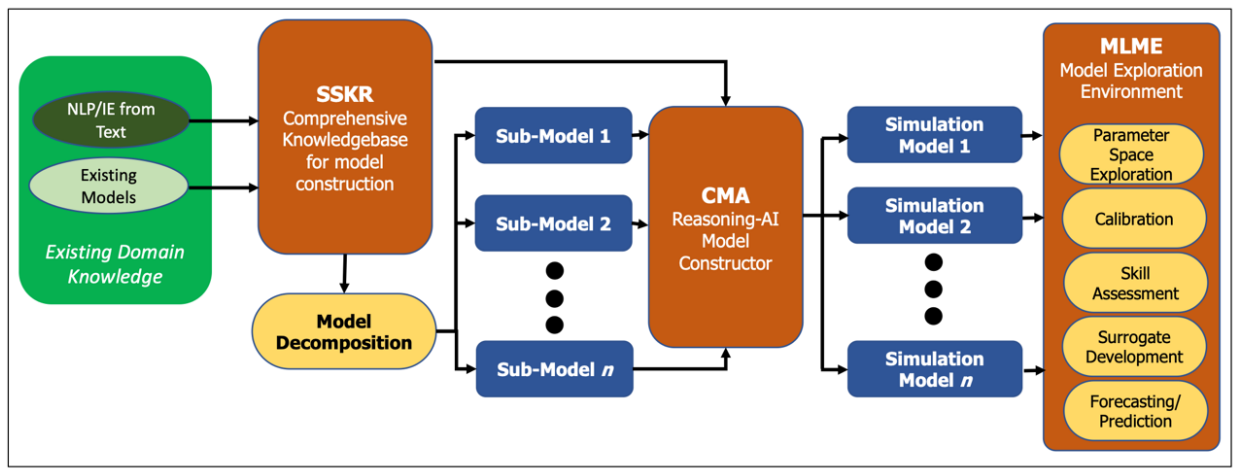

Figure 13. Overview of MAGCC: Information extracted (IE) from text via natural language processing (NLP) or existing computational models are inputted into the Structured Scientific Knowledge Representation (SSKR) object. Existing models can then be decomposed. The model representation is then fed into the Computational Modeling Assistant, which will determine which types of models or mathematical frameworks (e.g., ABM or differential equation) can use the provided knowledge. The CMA then generates simulator code. The simulation models are then operated upon using the Machine Learning Model Exploration environment, which will perform tasks such as model calibration, parameter space exploration, surrogate development, or fitness-for-task assessment. Figure and Caption reproduced from Ref [60] under the Creative Commons License.

The MAGCC Framework includes the ability to leverage ongoing developments in natural language processing (NLP) and large language models (LLMs) to extract biological knowledge from prior work, both textual literature and existing computational models, and through a series of steps, formally structure that knowledge to facilitate the creation of computational models (an extended MRM), instantiate those models into executable code, and the perform model exploration at scale using a refined version of the Nested GA/AL pipeline described above. Interested readers are directed to Ref [60] for more details.

### 4.6 Conclusion

We pose that the CIDT schema is an exemplar of responding to and following the NASEM Guidance for developing MDTs, and that it does so by focuses on bridging the gap between why currently employed industrial DTs are effective and what needs to be developed to make this concept applicable to biomedicine. The primary barrier to the development of MDTs is the difference between biological systems and the physical/process systems for which the concept of a DT was developed, namely that currently deployed industrial DTs are applied to *engineered* systems. As a feature of being engineered systems, industrial applications:

- Have existing underlying computational specifications that are invariably dynamic models that can be simulated; there is not ambiguity as to how these systems are supposed to work. This is not the case in biology, where a state of perpetual epistemic uncertainty and incompleteness exists.

- Have clear and identifiable output metrics in the real-world, and a clear and direct relationship between those metrics and the overall system-level states of the real-world object. This is a consequence of industrial objects having been engineered in a process that has predefined failure points and operational constraints. This is not the case in biology, where there is nearly intractable uncertainty as to what the actual ground truth is regarding the distributions of values of measured metrics and their relationship to targeted system-level phenotypes.

These differences lead to a coalescence of the findings and conclusions present in the NASEM Report that we have chosen to focus on:

- The dynamic nature of DTs.
- The need for ongoing updating between DTs and their real-world twins.
- DT design based on fit-for-purpose.
- Validation and uncertainty assessment as a precondition for establishing trustworthiness.
- Control as a goal of a MDT.

While the respective findings in the NASEM Report are necessarily divided into distinct sections for presentation purposes, the fact is that operationally the topics are interconnected and directly influence each other. For example, fit-for-purpose is intimately tied to defining and establishing trustworthiness, which involves validation and uncertainty assessment, particularly if control is specified as a central purpose of the MDT. The discovery of effective therapies for a population is the identification of robust control provided in a fashion that is generalizable across the entire clinical population. We see the goal of MDTs exactly in line with our proposed Axioms of True Precision Medicine, and as those Axioms have driven the development of the methods and approaches presented here, we find that this schema, which integrates those approaches, is ideally suited to facilitating the ability of MDT technology to achieve its full promise.

**Acknowledgements:**

This work was supported in part by the National Institutes of Health Award UO1EB025825. This research is sponsored by the Defense Advanced Research Projects Agency (DARPA) through Cooperative Agreement D20AC00002 awarded by the U.S. Department of the Interior (DOI), Interior Business Center. The content of the information does not necessarily reflect the position or the policy of the Government, and no official endorsement should be inferred.